\documentclass[reqno]{amsart}
\usepackage{graphicx,amsmath}
\usepackage{float}
\usepackage[utf8]{inputenc}
\usepackage{color}
\usepackage{cite}
\usepackage{fancyhdr}
\usepackage[margin=2cm]{geometry}
\usepackage{appendix}
\usepackage{multirow}
{ 

\begin{document}
\title{Novel evaluation method for non-Fourier effects in heat pulse experiments}

\author{A. Fehér$^{1}$, R. Kovács$^{123}$}

\address{
$^1$Department of Energy Engineering, Faculty of Mechanical Engineering, BME, Budapest, Hungary
$^2$Department of Theoretical Physics, Wigner Research Centre for Physics,
Institute for Particle and Nuclear Physics, Budapest, Hungary
$^3$Montavid Thermodynamic Research Group
}

\date{\today}

\begin{abstract}
The heat pulse (flash) experiment is a well-known and widely accepted method to measure the thermal diffusivity of a material. In recent years, it is observed that the thermal behavior of heterogeneous materials can show deviation from the classical Fourier equation, resulting in a different thermal diffusivity and requiring further thermal parameters to identify. Such heterogeneity can be inclusions in metal foams, layered structure in composites, or even cracks and porous parts in rocks. Furthermore, the next candidate, the so-called Guyer-Krumhansl equation, is tested on these experiments with success. However, these recent evaluations required a computationally intensive fitting procedure using countless numerical solutions, even when a good initial guess for the parameters is found by hand. This paper presents a Galerkin-type discretization for the Guyer-Krumhansl equation, which helped us find a reasonably simple analytical solution for time-dependent boundary conditions. Utilizing this analytical solution, we developed a new evaluation technique to immediately estimate all the necessary thermal parameters using the measured temperature history.

\end{abstract}
\maketitle

\section{Introduction}
\begin{figure}[H]
\centering
\includegraphics[width=10cm,height=7cm]{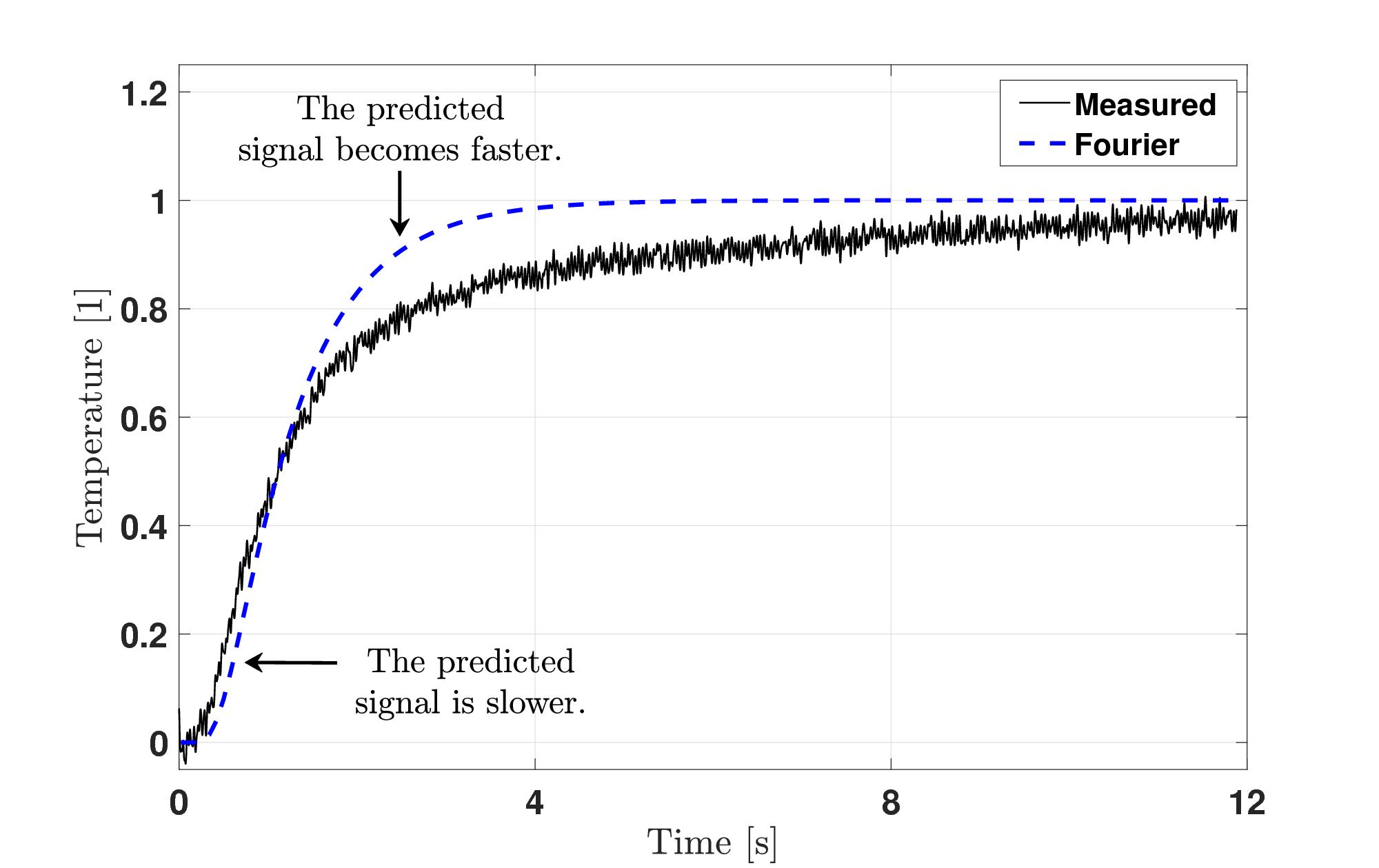}
\caption{Measured rear side temperature history for the capacitor sample and the prediction provided by Fourier's theory.}
 \label{fig5F}
\end{figure}

The engineering practice requires reliable ways to determine the necessary parameters, which are enough to characterize the material behavior. In what follows, we place our focus on the thermal description of materials, especially on heterogeneous materials such as rocks and foams. In recent papers \cite{Botetal16, Vanetal17}, it is reported that the presence of various heterogeneities can result in a non-Fourier heat conduction effect on macro-scale under room temperature conditions. A particular one is depicted in Fig.~\ref{fig5F} for a capacitor sample having a periodic layered structure. Such effects are observed in a so-called flash (or heat pulse) experiment in which the front side of the specimen is excited with a short heat pulse, and the temperature is measured at the rear side. That temperature history is used to find the thermal diffusivity in order to characterize the transient material behavior.

This non-Fourier effect occurs on a specific time interval as Fig.~\ref{fig5F} shows for a typical outcome of the flash experiments; this is called over-diffusion. After that interval, the Fourier equation appears to be a suitable choice for modeling, the influence of the heterogeneities vanishes (later, we show further examples). Also, there is no difference between the steady-states of Fourier and non-Fourier heat equations. Based on our experimental experience, the existence of over-diffusion depends on various factors, for instance, sample thickness, characteristic parallel time scales, and excitation (i.e., boundary conditions) \cite{FulEtal18e}.

In the following sections, we organize the discussion as follows. First, we briefly introduce the two heat conduction models to model the heat pulse experiments and used for evaluations with a particular set of dimensionless quantities. Second, we shortly present how the complete evaluation with the Fourier heat equation can be conducted. Then, we move on the way of the evaluation procedure with the Guyer-Krumhansl equation. After, we demonstrate the benefits of this fitting procedure and revisit some previous measurements. Furthermore, we decided to place the derivation of the analytical solutions to the end of the paper as an Appendix. According to our knowledge, the Galerkin method has not been used before for the Guyer-Krumhansl equation, and is a novel result in this respect, we want to keep the focus on its practical utilizations.

\section{Models for heat pulse experiments}
Although numerous generalizations of Fourier's law exist in the literature \cite{Van16}, there is solely one of them, which indeed proved to be reasonable as the next candidate beyond Fourier's theory, this is called Guyer-Krumhansl (GK) equation, this constitutive equation reads in one spatial dimension
\begin{align}
\tau_q \partial_t q + q+ \lambda \partial_x T - \kappa^2 \partial_{xx} q=0. \label{gk1}
\end{align}
Here, $\tau_q$ is the relaxation time for the heat flux $q$ and $\kappa^2$ is a kind of `dissipation parameter', usually related to the mean free path.
Whereas it was first derived on the basis of kinetic theory \cite{GuyKru66a1}, this model also has a strong background in non-equilibrium thermodynamics with internal variables (NET-IV) \cite{Van01a, VanFul12}. While in the first case, one assumes an underlying mechanism for phonons as the kinetic theory requires it, this is entirely neglected in the case of NET-IV, leaving the coefficients to be free (however, their sign is restricted by the II.~law of thermodynamics). Eq.~\eqref{gk1} is a time evolution equation for the heat flux, and in order to have a mathematically and physically complete system, we need the balance of internal energy $e$, too,
\begin{align}
\rho c \partial_t T + \partial_x q = 0, \label{bale}
\end{align}
in which the equation of state $e=cT$ is used with $c$ being the specific heat and $rho$ is the mass density. All these coefficients are constant, only rigid bodies are assumed with no heat source.

At this point, we owe an explanation of why we leave the Maxwell-Cattaneo-Vernotte (MCV) equation out of sight.
\begin{enumerate}
\item Hyperbolicity vs.~parabolicity. It is usually claimed that a heat equation model should be hyperbolic such as the MCV theory, describing finite propagation speed. Indeed, this seems reasonable, but it does not help in the practical applications under common conditions (room temperature, heterogeneous materials). The Fourier equation is still well-applicable in spite of its parabolic nature, therefore we do not see it as a decisive property.
\item In a low-temperature situation, the MCV model was useful, primarily due to the observed wave phenomenon in super-fluids, called second sound \cite{Tisza38, Lan47}. Despite the GK equation's parabolic nature, it also helped the researchers find the second sound in solids as well \cite{GK66}.
\item There is a significant effort to find the trace of wave propagation at room temperature (in a macro-scale object, so nano-structures does not count now), sadly with no success \cite{MitEta95, JozsKov20b}.
\item There are higher-order models as well, such as ballistic-diffusive models \cite{KovVan15, DreStr93a, MulRug98}, but they are related to a different research program, and for this work, investigating macro-scale objects, they do not seem relevant.
\item On the analogy of the MCV model, the so-called dual-phase lag (DPL) equation \cite{Tzou14} usually used in many works as the best candidate after Fourier's law. Sadly, this model introduces two time constants in an ad hoc manner, violating basic physical principles \cite{Ruk14, Ruk17}, leading to mathematically ill-posed problems as well \cite{FabEtal16, Fabetal14}.
\end{enumerate}

Last but not least, we also must mention a relatively less-known model from the literature, the Nyíri equation \cite{Nyiri91},
\begin{align}
q+ \lambda \partial_x T - \kappa^2 \partial_{xx} q=0,
\end{align}
which one is indeed similar to the Guyer-Krumhansl model but leaves the time lagging effects out of sight, hence it is purely a spatially nonlocal heat equation. Testing its solutions with the method presented in the Appendix, it turned out to be inaccurate for measurements, unfortunately. Consequently, the GK model is indeed the simplest but necessary extension for the Fourier equation, neither the MCV nor the Nyíri models are capable of describing these experiments accurately. In other words, the two new parameters ($\tau_q$ and $\kappa^2$) are truly needed.

\subsection{T and q-representations}
Depending on the purpose, it is useful to keep in mind that for such linear models, it is possible to chose a `primary' field variable, which could ease the definition of boundary conditions in some cases. For the GK equation, the temperature $T$ and the heat flux $q$ are the candidates, and their forms are
\begin{align}
\textrm{T-representation:}& \quad \tau_q \partial_{tt} T + \partial_t T - \alpha \partial_{xx} T - \kappa^2 \partial_{txx} T = 0, \\
\textrm{q-representation:}& \quad \tau_q \partial_{tt} q + \partial_t q - \alpha \partial_{xx} q - \kappa^2 \partial_{txx} q = 0.
\end{align}
We note that in $T$-representation, it is unknown how to define boundary condition for $q$ since it requires knowledge on $\partial_{xx} q$. On the other hand, in $q$-representation, it becomes meaningless to speak about $T$-boundaries. In a previous analytical solution for the GK equation \cite{Kov18gk}, this difference was inevitable to realize. In the present work, we use the system \eqref{gk1}-\eqref{bale}.
It is also interesting to notice that the GK model can recover the solution of the Fourier equation when $\kappa^2/\tau_q=\alpha$, this is called Fourier resonance \cite{Botetal16, VanKovFul15}. Overall, the coefficients $\tau_q$, $\alpha$, and $\kappa^2$ must be fitted to the given temperature history.

\subsection{Dimensionless set of parameters}
Following \cite{Botetal16}, we introduce these definitions for the dimensionless parameters (quantities with hat):
\begin{align}
&\textrm{time:} & \hat{t} =\frac{t}{t_p} \quad &  \textrm{and}  \quad \hat{x}=\frac{x}{L}; \nonumber \\
&\textrm{thermal diffusivity:} &  \hat \alpha = \frac{\alpha t_p}{L^2} \quad &\textrm{with} \quad \alpha=\frac{\lambda}{\rho c}; \nonumber \\
&\textrm{temperature:} & \hat{T}=\frac{T-T_{0}}{T_{\textrm{end}}-T_{0}} \quad &\textrm{with}\quad
T_{\textrm{end}}=T_{0}+\frac{\bar{q}_0 t_p}{\rho c L}; \nonumber \\
&\textrm{heat flux:} &
\hat{q}=\frac{q}{\bar{q}_0} \quad &\textrm{with}\quad
\bar{q}_0=\frac{1}{t_p}  \int_{0}^{t_p} q_{0}(t)\textrm{d}t; \nonumber \\
&\textrm{heat transfer coefficient:} & \hat h= h \frac{t_p}{\rho c}; \label{ndvar}
\end{align}
together with $\hat{\tau}_q =\frac{ \tau_q}{t_p}$,  $\hat{\kappa}^2 = \frac{\kappa^2}{L^2}$, where $\hat t$ differs from the usual Fourier number in order to decouple the thermal diffusivity from the time scale in the fitting procedure. Furthermore, $t_p$ denotes the constant heat pulse duration for which interval $\bar q_0$ averages the heat transferred with the heat pulse defined by $q_0(t)$. Here, $L$ is equal with the sample thickness. $T_{\textrm{end}}$ represents the adiabatic steady-state, and $T_0$ is the uniform initial temperature. In the rest of the paper, we shall omit the hat notation, otherwise we add the unit for the corresponding quantity. Utilizing this set of definitions, one obtains the dimensionless GK model:
\begin{align}
\partial_t T + \partial_x q&=0, \nonumber \\
\tau_q \partial_t q + q + \alpha \partial_x T -\kappa^2 \partial_{xx} q &= 0.
\end{align}
The initial condition is zero for both fields. For further details, we refer to the Appendix in which we present the analytical solution for the two heat equations. This set of dimensionless parameters does not change the definition of the Fourier resonance condition, i.e., it remains $\hat \kappa^2/\hat \tau_q = \hat \alpha$.

\section{Evaluation with the Fourier theory}
\begin{figure}[H]
\centering
\includegraphics[width=12cm,height=8cm]{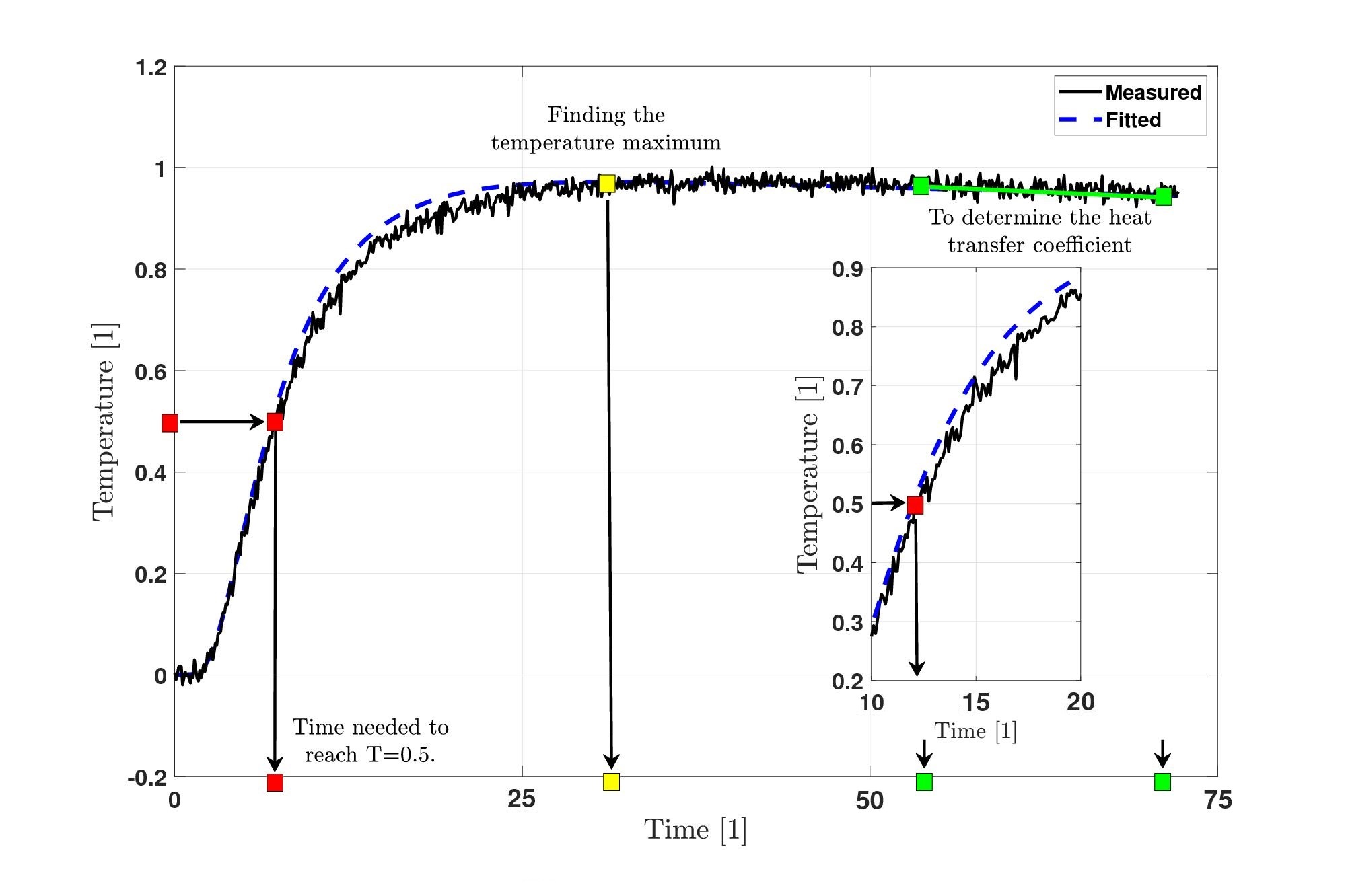}
\caption{Schematically presenting the evaluation method using Fourier's theory.}
 \label{fig1}
\end{figure}

The analytical solution of the Fourier equation is found for the rear side in the form of
\begin{align}
T(x=1,t) =Y_0 \exp(- h t) - Y_1 \exp(x_F t), \quad x_F=-2 h  - \alpha \pi^2, \quad t>30, \label{fouX1}
\end{align}
where all the coefficients are expressed in detail in the Appendix. First, we must estimate the heat transfer coefficient $h$ by choosing arbitrarily two temperature values at the decreasing part of temperature history. In this region, $\exp(x_F t )\approx 0$, thus
\begin{align}
h = - \frac{\ln(T_2/T_1)}{t_2 - t_1}.
\end{align}
For the Fourier theory, it is possible to express the thermal diffusivity explicitly, i.e.,
\begin{align}
\alpha_F= 1.38 \frac{L^2}{\pi^2 t_{1/2}}, \label{foueval}
\end{align}
and after registering $t_{1/2}$, it can be directly determined. This is the ratio of the thermal conductivity $\lambda$ and the specific heat capacity $\rho c$. Then, the top of the temperature history ($T_{\textrm{max}}$) follows by reading the time instant ($t_{\textrm{max}}$) when $T_{\textrm{max}}$ occurs. Figure \ref{fig1} schematically summarizes this procedure. Overall, we obtained the heat transfer coefficients, the thermal diffusivity and $T_{\textrm{max}}$, which all used for the Guyer-Krumhansl theory.

\section{Evaluation with the Guyer-Krumhansl theory}
The situation here becomes more difficult since this non-Fourier theory consists of two `time constants' ($x_1$ and $x_2$) instead of one ($x_F$ in the Fourier theory). Consequently, it is not possible to find these exponents without making simplifications, in which one must be immensely careful. We prepared `parameter maps' for all possible $\tau_q$ and $\kappa^2$ values that could be practically possible and beyond in order to check the effect of the simplifications made in the following. However, we still had to restrict ourselves to a domain, which is $3>\kappa^2/(\alpha \tau_q)\geq 1$. Its lower limit expresses the Fourier case, and any other combination falls on the over-diffusive region. The highest experimentally observed ratio so far is around $2.5$, thus we expect $3$ to be eligible. For $\kappa^2$, we consider $0.02<\kappa^2<1$.
We want to emphasize that the GK theory itself is not restricted on this domain, it would allow under-diffusive (`wave-like') propagation as well \cite{GK66}. However, for the present situation, we consider it out of interest in the lack of experimental observation for room temperature experiments on macro-scale heterogeneous samples.
In the GK theory, we can express the rear side temperature history as
\begin{align}
T(x=1,t) =Y_0 \exp(- h t) - Z_1 \exp(x_1 t) - Z_2 \exp(x_2 t), \quad x_1, x_2<0,
\end{align}
for the detailed calculation and parameter definitions, we refer to the Appendix again. This can be equivalently formulated realizing that $Z_2 = -P_0 - Z_1$,
\begin{align}
T(x=1,t) =Y_0 \exp(- h t) - Z_1 \big(\exp(x_1 t) - \exp(x_2 t) \big) + P_0  \exp(x_2 t),
\end{align}
where merely one simplification becomes possible for all $\tau_q$ and $\kappa^2$: $\exp(x_1 t)\gg \exp(x_2 t)$ when $t>60$, i.e.,
\begin{align}
T(x=1,t>60) =Y_0 \exp(- h t) - Z_1\exp(x_1 t)  + P_0  \exp(x_2 t).
\end{align}
This form is more advantageous because $P_0$ remains practically constant for a given boundary condition, thus its value can be assumed a-priori, this is exploited in the evaluation method. Now, let us present from step by step the determination of GK parameters, depicted on Fig.~\ref{fig2}.
\begin{itemize}
\item Step 1/A. We have to observe that the temperature predicted by Fourier's theory always runs together with the measured one at the beginning, after that, it rises faster at the top. In other words, in this region the same temperature value (usually around $0.7-0.95$) is reached sooner. Mathematically, we can express it by formally writing the equations for the Fourier and GK theories as follows,
\begin{align}
T_F = Y_0 \exp(-h t) - Y_1\exp(x_F t_F); \quad T_{GK}=Y_0 \exp(-h t_m) - Z_1\exp(x_1 t_m)  + P_0  \exp(x_2 t_m),
\end{align}
where the $t_F$ time instant is smaller than the measured $t_m$, also $T_F=T_{GK}$ holds. Let us choose such two temperatures arbitrarily and taking their ratio, it yields
\begin{align}
\exp\big(x_F (t_{F1} - t_{F2})\big) = \exp \big(x_1 (t_{m1}-t_{m2}) \big) \frac{-Z_1 + P_0 \exp\big((x_2 - x_1) t_{m1}\big)}{-Z_1 + P_0 \exp\big((x_2 - x_1) t_{m2}\big)}
\end{align}
where the fraction on the right hand side is close to $1$, mostly between $1$ and $1.05$ for `small' time intervals. It could be possible to introduce it as a correction factor (denoted with $c$ below) for $x_1$ in an iterative procedure if more to be known about $\tau_q$ and $\kappa^2$.
After rearrangement, we obtain a closed form formula for $x_1$:
\begin{align}
x_1 = \frac{\ln(1/c)}{t_{m1}-t_{m2}} + x_F \frac{t_{F1} - t_{F2}}{t_{m1}-t_{m2}}.
\end{align}
Taking $c=1$ is equivalent with neglecting $\exp(x_2 t)$ from the beginning around reaching $T_{\textrm{max}}$, and leading to this same expression. Eventually, it introduces a correction for the Fourier exponent $x_F$ based on the deviation from the measured data with the possibility to apply further corrections using $c$ if needed. Practically, we take the $80$\% and $90$\% of $T_{\textrm{max}}$ and  for the next $20$ subsequent measurement points, then we consider their mean value to be $x_1$. From mathematical point of view, closer data point pairs should perform better, but it does not due to the uncertainty in the measurement data. According to our experience, it offers a more consistent value for $x_1$.

\begin{figure}[H]
\centering
\includegraphics[width=11cm,height=8cm]{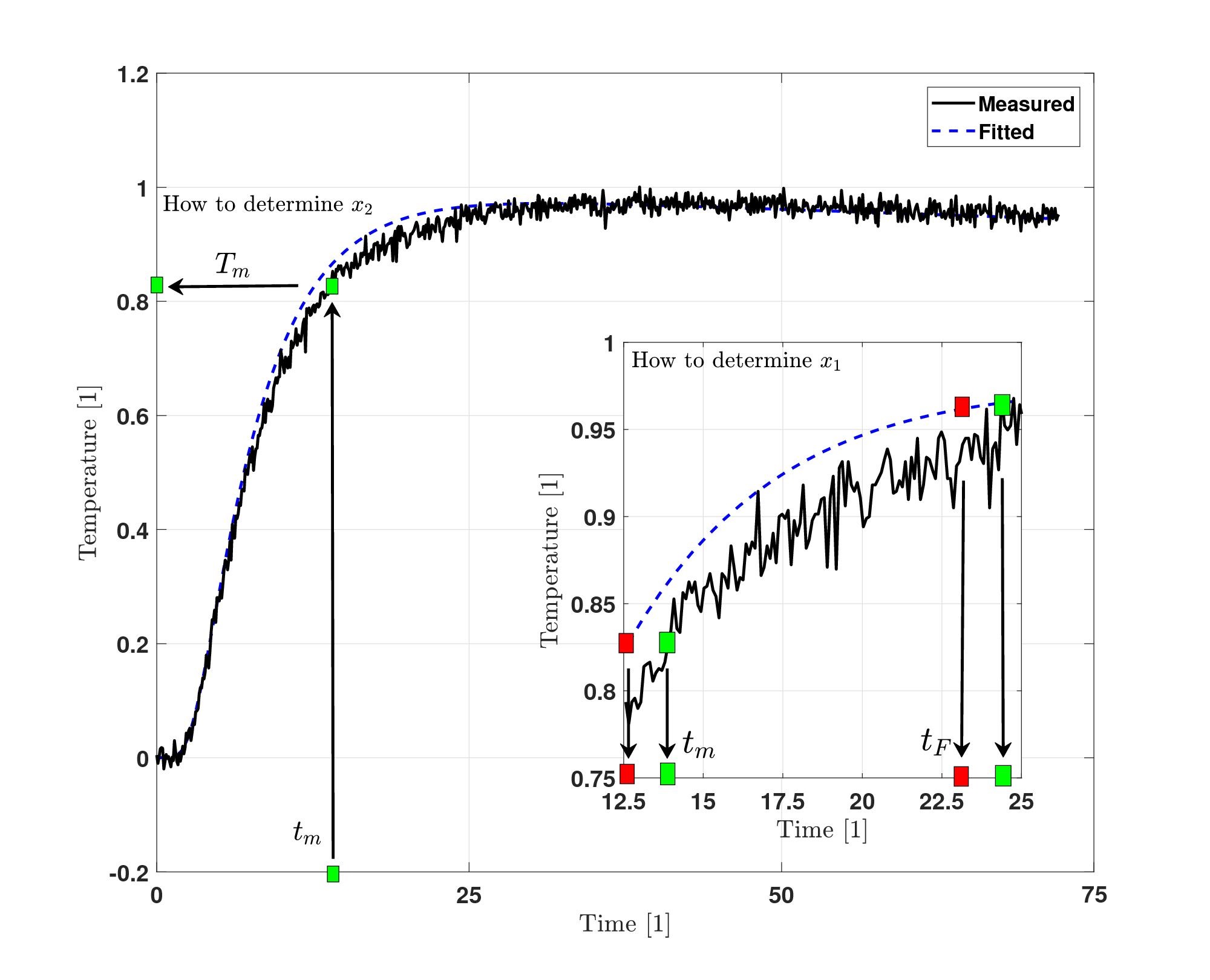}
\caption{The schematic representation of the evaluation method using the Guyer-Krumhansl theory. Here, the 'fitted curve' belongs to the Fourier equation.}
 \label{fig2}
\end{figure}

\item Step 1/B. In parallel with part A, we can determine the coefficient $Z_1$ for each $t_m$ and for each corresponding $x_{1,m}$, that is,
\begin{align}
Z_{1,m} = \exp(-x_{1,m} t_m) \Big(T_m - Y_0 \exp(-h t_m)\Big)
\end{align}
where the subscript $m$ denotes the value related to one measurement point. Also, after $20$ subsequent points, we take the mean value of the set $\{Z_{1,m}\}$.

\item Step 2. At this point, we can exploit that $P_0$ is `almost constant', i.e., $2<-P_0<2.03$ holds. Here, $2.03$ comes from the parameter sweep, we did not observe higher values for $-P_0$, and also, it cannot be smaller than $2$. This property allows us to a-priori assume its value (such as $P_0=-2.015$), and in a later step, we must fine-tune since the overall outcome reacts sensitively. Using $P_0$, we can obtain $Z_2=-P_0 - Z_1$. In order to obtain $x_2$, we can rearrange the equation
\begin{align}
T=Y_0 \exp(- h t) - Z_1 \exp(x_1 t) - Z_2 \exp(x_2 t)
\end{align}
for $x_2$, and calculate it as a mean value of the set $\{x_{2,m}\}$ filled with values related to each $t_m$. When having noisy data, this approach can result in positive $x_2$ values, unfortunately. These values must be excluded, otherwise, it leads to instability and a meaningless outcome. Careful data filtering can help to solve this shortcoming, and in fact, we used it to ease the calculation (the details are plotted in the next section).

\item Step 3. Now, having both exponents and coefficients, it is possible to rearrange the analytical expressions to the GK parameters explicitly and calculate $\alpha_{GK}$, $\tau_q$ and $\kappa^2$:
\begin{align}
x_1, x_2 \Rightarrow k_1, k_2; \quad Z_1 \Rightarrow DP_0 \Rightarrow M_1, M_2 \Rightarrow \tau_q \Rightarrow \alpha_{GK} \Rightarrow \kappa^2.
\end{align}
For the detailed parameter definitions, we refer to the Appendix.
\item Step 4. As it is mentioned in Step 2, the overall outcome is sensitive to $P_0$. Therefore we choose to make a sweep on the possible interval with the step of $0.002$, producing the temperature history for each set of parameters and characterizing them with $R^2$, the coefficient of determination. Lastly, we chose the best set.
\end{itemize}
Practically, this evaluation method reduces the number of `fitted' parameters as only $P_0$ has to be fine-tuned at the end. Besides, it is constrained into a relatively narrow range, consequently, the overall evaluation procedure takes only a few seconds instead of hours to perform computationally intensive algorithms.

\section{Comparison with foregoing experiments}
\begin{figure}[H]
\centering
\includegraphics[width=12cm,height=4cm]{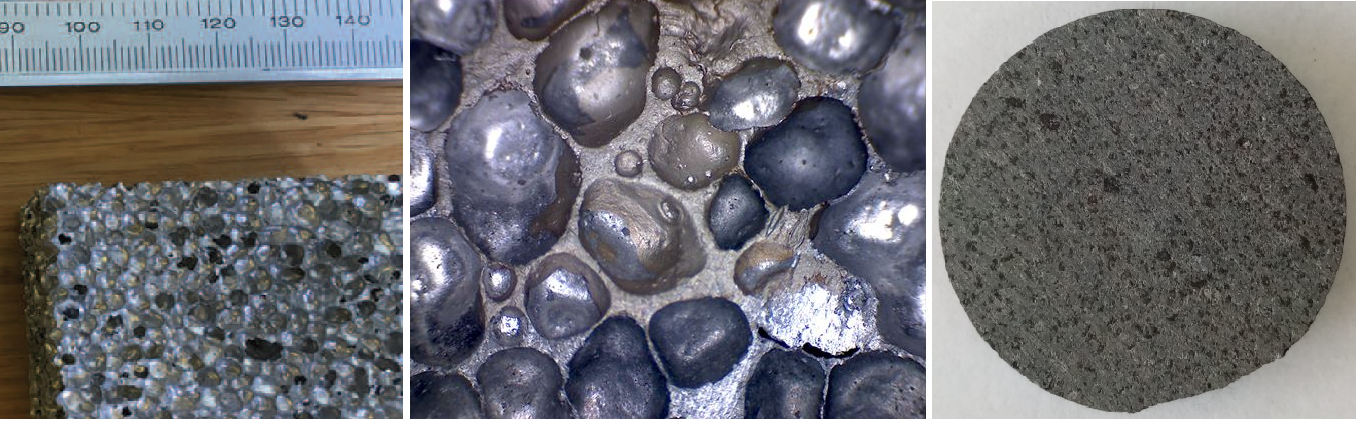}
\caption{The magnified view of the metal foam specimen (center) and the basalt rock sample (right).}
 \label{fig9}
\end{figure}

First, we revisit the experiments presented in \cite{FulEtal18e} since that set of data on Basalt rock samples with thicknesses of $1.86$, $2.75$ and $3.84$ mm, showed size dependence both on the thermal diffusivity and on the non-Fourier effects. However, the fitted parameters in \cite{FulEtal18e} found by hand, thus not exactly precise. Here, we aim to specify the exact quantities for the GK model and establishing a more robust theoretical basis for the observations. Second, we reevaluate the data recorded on a metal foam sample with $5.2$ mm thickness. This belongs to the samples showing the potent non-Fourier effect, presented first in \cite{Vanetal17}. Figure \ref{fig9} shows these samples.

\begin{figure}[H]
\centering
\includegraphics[width=13cm,height=5.6cm]{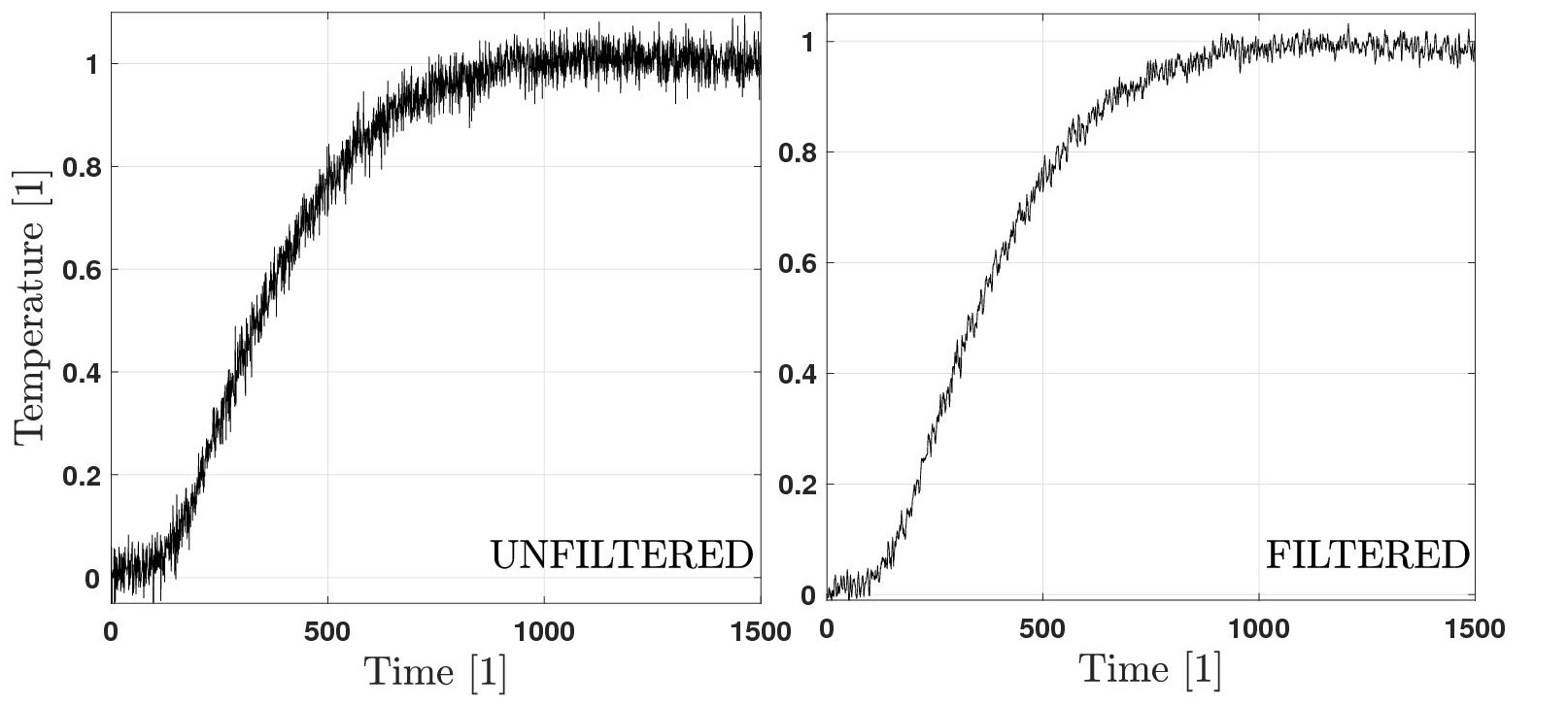}
\caption{The effect of data smoothening, using $10$ neighboring points in the Savitzky-Golay algorithm.}
 \label{fig8C}
\end{figure}

In some cases, the available data is too noisy for such an evaluation method, an example is presented in Fig.~\ref{fig8C}. That data is smoothed using the built-in Savitzky-Golay algorithm of Matlab. Besides, we paied much attention to not smooth it overly in order to keep the physical content untouched as much as possible.

\subsection{Basalt rock samples}
\begin{figure}[H]
\centering
\includegraphics[width=16cm,height=6.5cm]{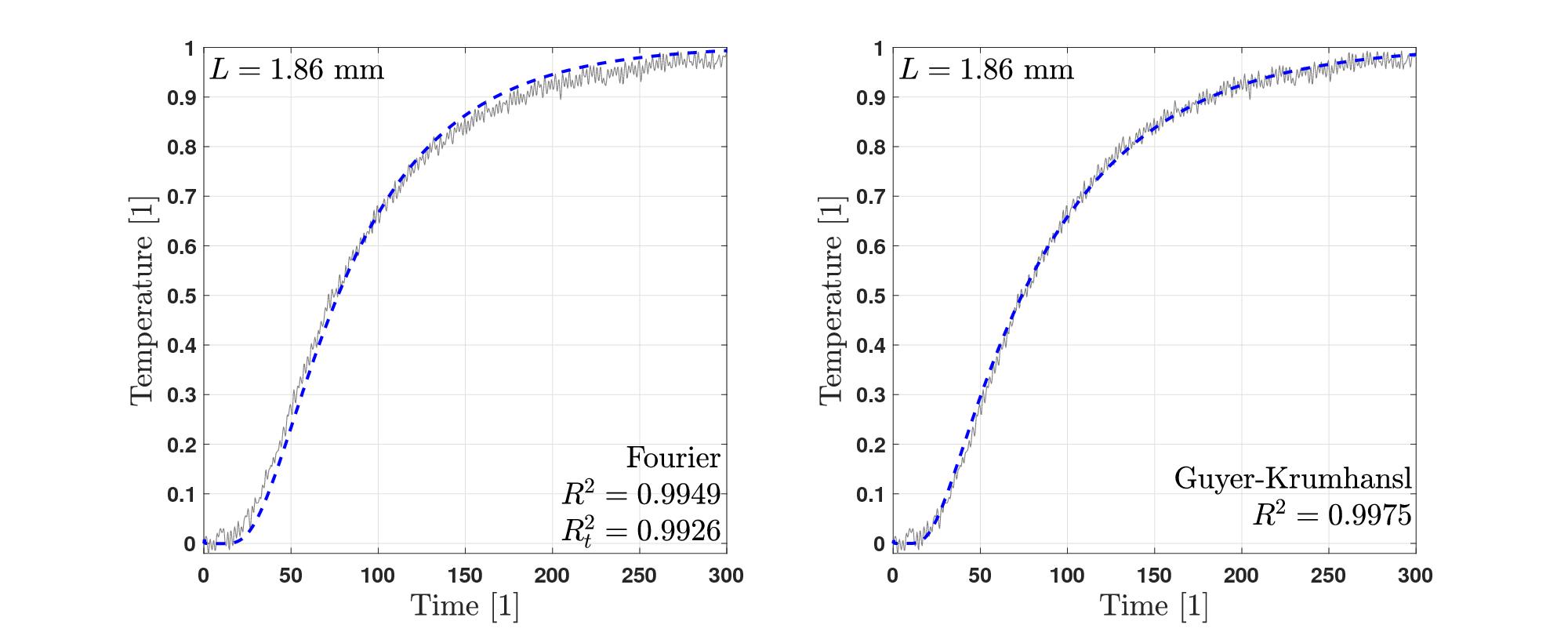}
\caption{The rear side temperature history for the basalt rock sample with $L=1.86$ mm.}
 \label{fig10B1A}
\end{figure}

\begin{figure}[H]
\centering
\includegraphics[width=16cm,height=6.5cm]{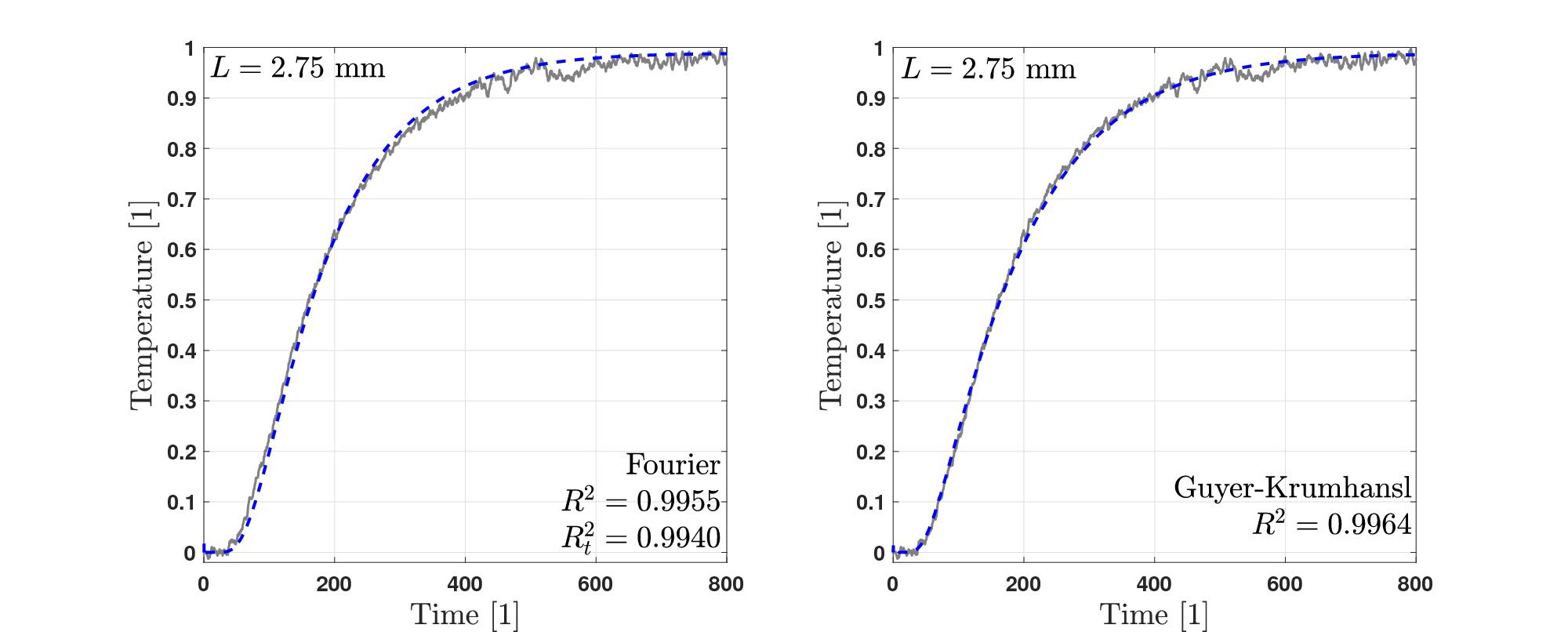}
\caption{The rear side temperature history for the basalt rock sample with $L=2.75$ mm.}
 \label{fig11B2A}
\end{figure}

\begin{figure}[H]
\centering
\includegraphics[width=12cm,height=6.5cm]{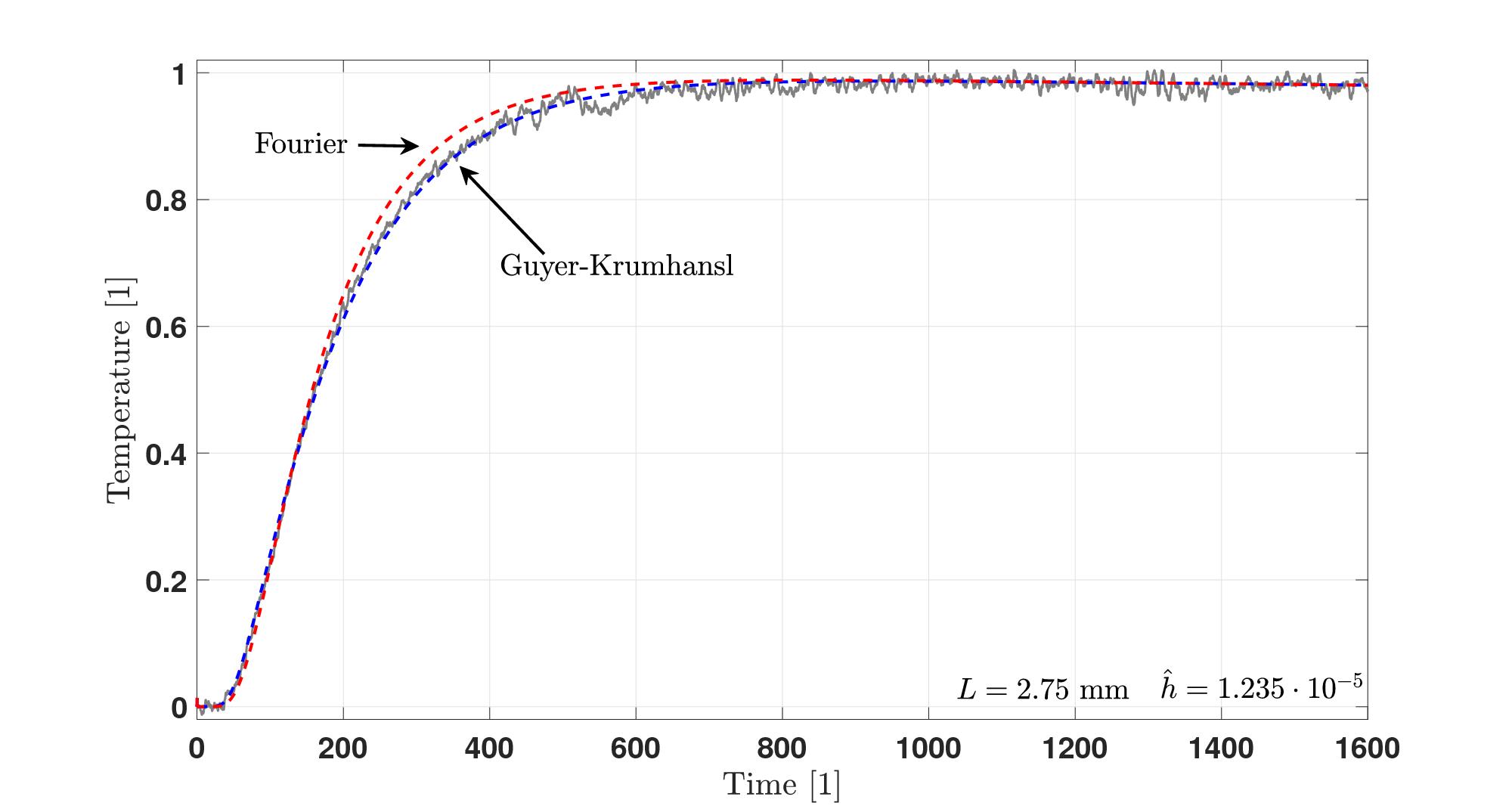}
\caption{Demonstrating the complete fitting for the rear side temperature in case of the basalt rock sample with $L=2.75$ mm.}
 \label{fig14B2B}
\end{figure}

\begin{figure}[H]
\centering
\includegraphics[width=16cm,height=6.5cm]{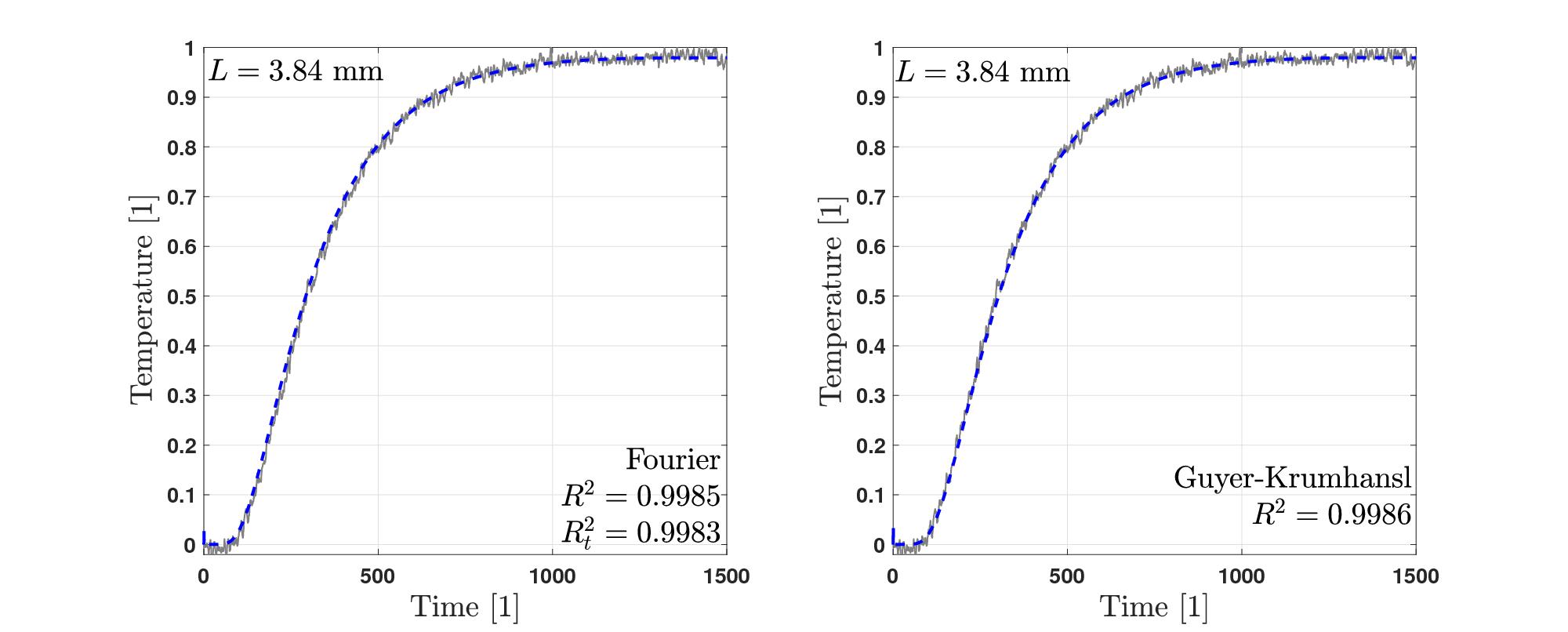}
\caption{The rear side temperature history for the basalt rock sample with $L=3.84$ mm.}
 \label{fig12B3A}
\end{figure}

Regarding the exact details of measurements, we refer to \cite{FulEtal18e}. Tables \ref{bazalt1} and \ref{bazalt2} consist of our findings using this evaluation algorithm. Comparing the outcomes of the two fitting procedures, we find the thermal diffusivities to be close to each other. However, this is not the case with the GK parameters $\tau_q$ and $\kappa^2$, they significantly differ from the previous values from \cite{FulEtal18e}. Despite the huge difference, the size-dependence for both the Fourier and non-Fourier behaviors is apparent nevertheless. The fitted temperature histories are depicted in Figs.~\ref{fig10B1A}, \ref{fig11B2A}, \ref{fig14B2B} and \ref{fig12B3A} for each thickness, respectively. Each figure shows the $R^2$ value for the fitted curve. For the Fourier one, two of them are given: $R_t^2$ represents the one found without any fine-tuned thermal diffusivity, this is purely theoretical. The other $R^2$ stands for fine-tuned $\alpha_F$.

In the first case ($L=1.86$ mm), although the difference for the non-Fourier samples seems negligible, it results in $10$\% difference in the thermal diffusivity. It is more visible from Table \ref{bazalt2}, in which the Fourier resonance condition spectacularly characterize the deviation from Fourier's theory, it decreases for thicker samples. Regarding the third one ($L=3.84$ mm), Fourier's theory seems to be `perfectly splendid', and the GK model hardly improves it. Indeed, the $0.94$ for the Fourier resonance is close enough to $1$ to consider it to be a Fourier-like propagation.

\begin{table}[H]
\begin{tabular}{c|c|c|c|c|cccc}
\multirow{2}{*}{\begin{tabular}[c]{@{}c@{}}Basalt rock \\ samples\end{tabular}} & \multicolumn{4}{c|}{Findings in \cite{FulEtal18e}}                                                                                                                                                                                                                                                  & \multicolumn{4}{c}{Refined results}                                                                                                                                                                                                                                                                                                                                       \\ \cline{2-9}
                                                                                & \begin{tabular}[c]{@{}c@{}}$\alpha_F$\\  $10^{-6}$ {[}m$^2$/s{]}\end{tabular} & \begin{tabular}[c]{@{}c@{}}$\alpha_{GK}$\\  $10^{-6}$ {[}m$^2$/s{]}\end{tabular} & \begin{tabular}[c]{@{}c@{}}$\tau_q$\\ {[}s{]}\end{tabular} & \begin{tabular}[c]{@{}c@{}}$\kappa^2$\\ $10^{-6}$ {[}m$^2${]}\end{tabular} & \multicolumn{1}{c|}{\begin{tabular}[c]{@{}c@{}}$\alpha_F$\\  $10^{-6}$ {[}m$^2$/s{]}\end{tabular}} & \multicolumn{1}{c|}{\begin{tabular}[c]{@{}c@{}}$\alpha_{GK}$\\  $10^{-6}$ {[}m$^2$/s{]}\end{tabular}} & \multicolumn{1}{c|}{\begin{tabular}[c]{@{}c@{}}$\tau_q$\\ {[}s{]}\end{tabular}} & \begin{tabular}[c]{@{}c@{}}$\kappa^2$\\ $10^{-6}$ {[}m$^2${]}\end{tabular} \\ \hline
$1.86$ mm                                                                       & $0.62$                                                                        & $0.55$                                                                           & $0.738$                                                    & $0.509$                                                                    & \multicolumn{1}{c|}{$0.68$}                                                                        & \multicolumn{1}{c|}{$0.61$}                                                                           & \multicolumn{1}{c|}{$0.211$}                                                    & $0.168$                                                                    \\ \hline
$2.75$ mm                                                                       & $0.67$                                                                        & $0.604$                                                                          & $0.955$                                                    & $0.67$                                                                     & \multicolumn{1}{c|}{$0.66$}                                                                        & \multicolumn{1}{c|}{$0.61$}                                                                           & \multicolumn{1}{c|}{$0.344$}                                                    & $0.268$                                                                    \\ \hline
$3.84$ mm                                                                       & $0.685$                                                                       & $0.68$                                                                           & $0.664$                                                    & $0.48$                                                                     & \multicolumn{1}{c|}{$0.70$}                                                                        & \multicolumn{1}{c|}{$0.68$}                                                                           & \multicolumn{1}{c|}{$1$}                                                        & $0.65$
\end{tabular}
\caption{Summarizing the fitted thermal parameters.}
\label{bazalt1}
\end{table}

\begin{table}[H]
\begin{tabular}{c|c|c}
$\frac{\kappa^2}{\tau_q \alpha}$ & \begin{tabular}[c]{@{}c@{}}Findings \\ in \cite{FulEtal18e}\end{tabular} & \begin{tabular}[c]{@{}c@{}}Refined\\ values\end{tabular} \\ \hline
$1.86$ mm                        & $1.243$                                                        & $1.295$                                                  \\ \hline
$2.75$ mm                        & $1.171$                                                        & $1.272$                                                  \\ \hline
$3.84$ mm                        & $1.06$                                                         & $0.94$
\end{tabular}
\caption{Characterizing the non-Fourier behavior using the Fourier resonance condition.}
\label{bazalt2}
\end{table}

\subsection{Metal foam} 
\begin{figure}[H]
\centering
\includegraphics[width=16cm,height=6.5cm]{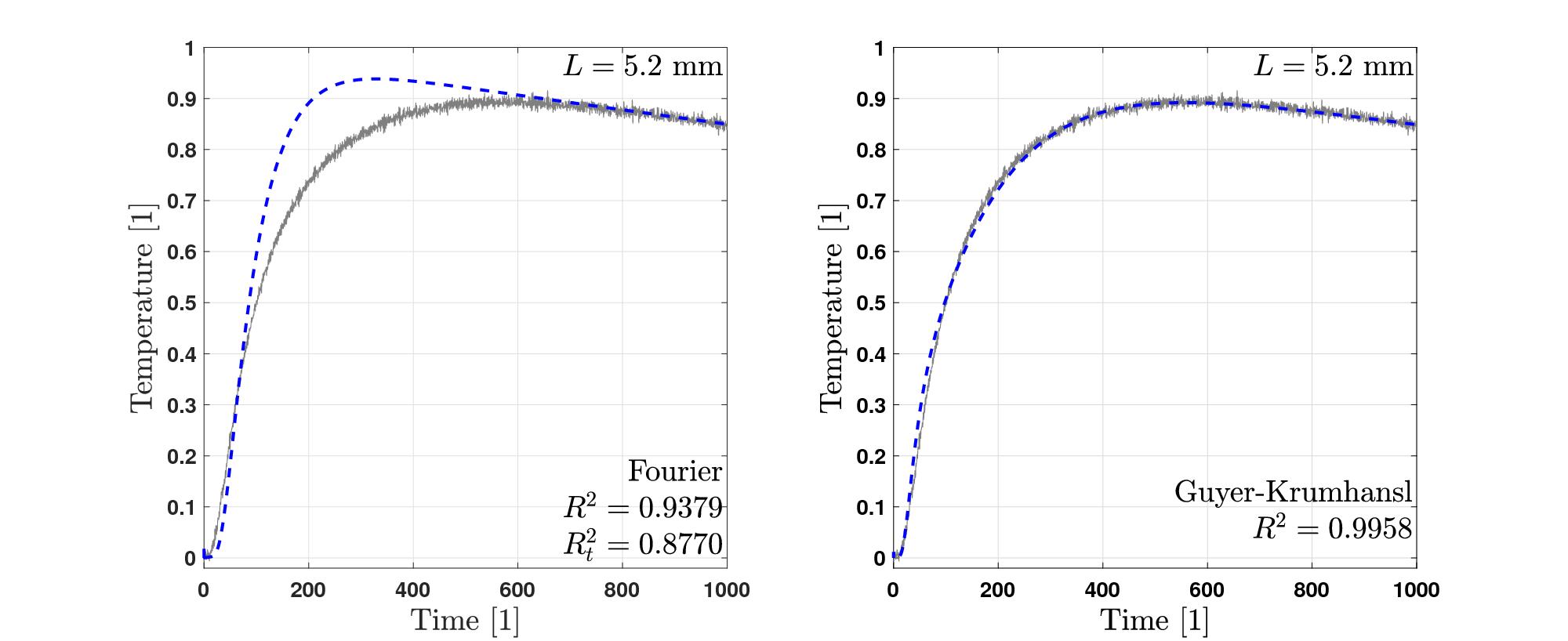}
\caption{The rear side temperature history for the metal foam sample with $L=5.2$ mm.}
 \label{fig13MF1A}
\end{figure}

Regarding the extent of the non-Fourier effect, the situation becomes remarkably different for the metal foam sample, presented first in \cite{Vanetal17}. The millimeter size inclusions can significantly influence the thermal behavior. The outcome is plotted in Fig.~\ref{fig13MF1A} together with the corresponding $R^2$ values. Table \ref{MF1} helps to compare the fitted values found by Wolfram Mathematica to ours. Notwithstanding that the GK parameters are in correspondence, the most notable difference is on the thermal diffusivities, interestingly. The Fourier resonance parameter is found to be $2.395$ with our procedure, while on the contrary to $3.04$ in \cite{Vanetal17}. Common in both cases, the ratio of $\alpha_F$ and $\alpha_{GK}$ is found to be $1.28$-$1.29$, which represents an indeed remarkable deviation from Fourier's theory.

\begin{table}[H]
\begin{tabular}{c|c|c|c|c|cccc}
\multirow{2}{*}{\begin{tabular}[c]{@{}c@{}}Metal foam \\ sample\end{tabular}} & \multicolumn{4}{c|}{Findings in \cite{Vanetal17} with Wolfram Math}                                                                                                                                                                                                                                                  & \multicolumn{4}{c}{Present algorithm}                                                                                                                                                                                                                                                                                                                                       \\ \cline{2-9}
                                                                                & \begin{tabular}[c]{@{}c@{}}$\alpha_F$\\  $10^{-6}$ {[}m$^2$/s{]}\end{tabular} & \begin{tabular}[c]{@{}c@{}}$\alpha_{GK}$\\  $10^{-6}$ {[}m$^2$/s{]}\end{tabular} & \begin{tabular}[c]{@{}c@{}}$\tau_q$\\ {[}s{]}\end{tabular} & \begin{tabular}[c]{@{}c@{}}$\kappa^2$\\ $10^{-6}$ {[}m$^2${]}\end{tabular} & \multicolumn{1}{c|}{\begin{tabular}[c]{@{}c@{}}$\alpha_F$\\  $10^{-6}$ {[}m$^2$/s{]}\end{tabular}} & \multicolumn{1}{c|}{\begin{tabular}[c]{@{}c@{}}$\alpha_{GK}$\\  $10^{-6}$ {[}m$^2$/s{]}\end{tabular}} & \multicolumn{1}{c|}{\begin{tabular}[c]{@{}c@{}}$\tau_q$\\ {[}s{]}\end{tabular}} & \begin{tabular}[c]{@{}c@{}}$\kappa^2$\\ $10^{-6}$ {[}m$^2${]}\end{tabular} \\ \hline
$5.2$ mm                                                                       & $3.04$                                                                        & $2.373$                                                                           & $0.402$                                                    & $2.89$                                                                    & \multicolumn{1}{c|}{$3.91$}                                                                        & \multicolumn{1}{c|}{$3.01$}                                                                           & \multicolumn{1}{c|}{$0.304$}                                                    & $2.203$
\end{tabular}
\caption{Summarizing the fitted thermal parameters for the metal foam sample.}
\label{MF1}
\end{table}

\section{Discussion and summary}
We developed an algorithm to efficiently evaluate room temperature heat pulse experiments in which a non-Fourier effect could exist. This is called over-diffusive propagation and detunes the thermal diffusivity, even when the deviation is seemingly small or negligible for the rear side temperature history. The presented method is based on the analytical solution of the Guyer-Krumhansl equation, including temperature-dependent convection boundary condition, thus the heat transfer to the environment can be immediately included in the analysis. The reevaluation of preceding experiments showed a real size-dependence for all thermal parameters, especially for the GK coefficients $\tau_q$ and $\kappa^2$. Furthermore, it is in accordance with the result of the iterative `brute force' iterative fitting procedure of Wolfram Math, basically, but using much less computational resource.

We plan to improve this procedure by including the investigation of front side temperature history, too. When $x_1$ is obtained from the rear side, it could be easily used to describe the front side's thermal behavior. This is much more sensitive to the initial time evolution right after the excitation, therefore it could serve as a better candidate to achieve a more precise and robust estimation for the $x_2$ exponent. Also, having two temperature histories would be a remarkable step forward to ascertain the existence of non-Fourier heat conduction.

We believe that this procedure lays the foundations for the more practical engineering applications of non-Fourier models, especially for the best candidate among all of them, the Guyer-Krumhansl equation. It sheds new light on the classical and well-known flash experiments, and we provide the necessary tools to find additional thermal parameters to achieve a better description of heterogeneous materials. It becomes increasingly important with the spreading of composites and foams and helps characterize 3D printed samples with various inclusions.

\section{Acknowledgement}
The authors thank Tamás Fülöp, Péter Ván, and Mátyás Szücs for the valuable discussions. We thank László Kovács (Kőmérő Kft., Hungary) and Tamás Bárczy (Admatis Kft.) for producing the rock and metal foam samples.

The research reported in this paper and carried out at BME has been supported by the grants National
Research, Development and Innovation Office-NKFIH FK 134277, and by the NRDI Fund (TKP2020 NC, Grant No. BME-NC) based on the charter of bolster issued by the NRDI Office under the auspices of the Ministry for Innovation and Technology.
This paper was supported by the János Bolyai Research Scholarship of the Hungarian Academy of Sciences.

\section{Appendix: Galerkin-type solution of heat equations}
Since the non-Fourier models are not well-known in the general literature, there are only a few available analytical and numerical methods to solve such a spatially nonlocal equation like the Guyer-Krumhansl one. The nonlocal property is a cornerstone of these models since the usual boundary conditions do not work in the same way. That could be a problem when the outcome seemingly violates the maximum principle, see for instance \cite{Zhukov16} in which the operational approach is applied \cite{Zhu16a}.

Another particular candidate originates from the spectral methods, it is called Galerkin method, where both the weight and the trial functions are the same. Fortunately, following \cite{Kov18gk}, we can surely apply sine and cosine trial functions in which terms the solution can be expressed. It is important to emphasize that we deal with a system of partial differential equations in our case. The physical (and mathematical) connection between the field variables restricts the family of trial functions. Namely, even in the simplest case of the Fourier heat equation,
\begin{align}
\partial_t T + \partial_x q=0, \quad  q + \alpha \partial_x T  = 0,
\end{align}
$q$ and $T$ are orthogonal to each other, and the trial functions must respect this property. Our choice is found by the method called separation of variables, but it resulted in a too complicated outcome due to the time-dependent boundary condition. In \cite{Kov18gk}, the heat pulse is modeled with a smooth $q_0(t)=1- \cos( 2 \pi t/t_p)$ function on the $0<t\leq t_p$ interval. It is disadvantageous since the most interesting part falls beyond $t_p$, and the solution in $t_p<t$ must account for the state at the time instant $t_p$ as an initial condition. Therefore, it results in cumbersome expressions for the coefficients.

We overcome this difficulty by introducing a different function to model the heat pulse, i.e., we use $q_0(t) = -(\exp(-C_1 t) +\exp(-C_2 t))/n$, where $C_1$ and $C_2$ are chosen to have a sufficiently small $q_0$ after $t_p$, hence the values are $C_1=1/0.075$ and $C_2=6$. The coefficient $n$ is normalizing $q_0$ to $1$ from $0$ to $t_p$, so it is $n=(C_1-C_2)/(C_1 C_2)$. For larger time instants, the front side becomes adiabatic. Regarding the rear side boundary condition, we choose to account heat convection for both models.
\begin{enumerate}
\item First, we restrict ourselves to the Fourier equation. According to our previous experiments \cite{Botetal16, Vanetal17, FulEtal18e}, one can safely use the Fourier equation where cooling effects become significant. This solution is used to estimate the heat transfer coefficient, the maximum temperature and give a first approximation to the thermal diffusivity.

\item In the second step, we repeat the calculation for the GK model with the same boundary conditions. We use the previously found Fourier parameters as the input to estimate the GK parameters and fine-tune the thermal diffusivity. The heat transfer coefficient and the temperature maximum can be kept the same.
\end{enumerate}

\subsection{Step 1: solving the Fourier equation}
\begin{figure}
\centering
\includegraphics[width=13cm,height=10cm]{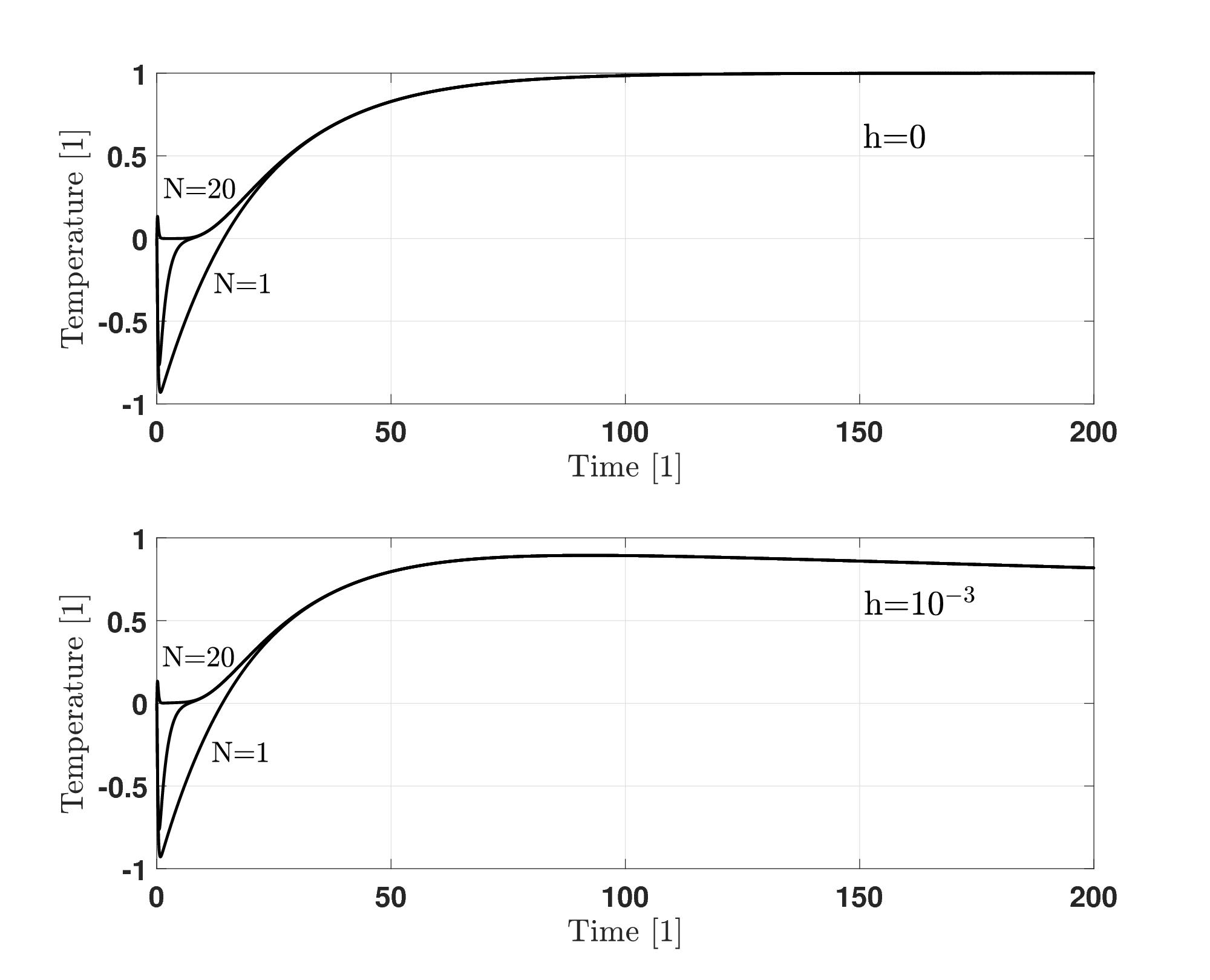}
\caption{Convergence analysis for the Fourier equation in two different cases on the rear side temperature history. The first one shows the adiabatic limit, and the second one presents the case when the heat transfer coefficient $h$ is not zero. In both cases, we applied $1$, $3$ and $20$ terms in the spectral decomposition \eqref{specFOU}, with $\alpha = 0.005$.}
 \label{fig3}
\end{figure}

While there are several available solutions in the literature, we want to see how the Galerkin approach performs on this model using our set of dimensionless parameters and boundary conditions. Consequently, we can keep our findings to be consistent  between the two heat equation models. Let us recall the mathematical model for the sake of traceability. In the Fourier model, we have
\begin{align}
\partial_t T + \partial_x q=0, \quad  q + \alpha \partial_x T = 0, \label{AFOU}
\end{align}
with
\begin{align}
q_0(t)= -\frac{1}{n}\Big(\exp(-C_1 t) +\exp(-C_2 t)\Big),\quad  n=\frac{C_1-C_2}{C_1 C_2}, \quad q_1(t) = h T(x=1,t),
\end{align}
in which all parameters are dimensionless as presented in Sec.~2.2. The initial conditions are $q(x,t=0)=0$ and $T(x,t=0)=0$, the conducting medium is thermally relaxed. We emphasize that one does not need to separately specify boundary conditions for the temperature field $T$ as well. Regarding the heat flux field $q$, we must separate the time-dependent part from the homogeneous one,
\begin{align}
q(x,t) = w(x,t) + \tilde q (x,t), \quad w(x,t)=q_0(t) + x \left (q_1(t) - q_0(t) \right)
\end{align}
with $\tilde q$ being the homogeneous field and $w$ inherits the entire time-dependent part from the boundary (and $x$ runs from $0$ to $1$).
The spectral decomposition of $\tilde q$ and $T$ are
\begin{align}
\tilde q(x,t)=\sum_{j=1}^{N} a_j (t) \phi_j (x), \quad T(x,t)=\sum_{j=1}^{N} b_j (t) \varphi_j (x), \label{specFOU}
\end{align}
with $\phi_j(x)=\sin(j \pi x)$ and $\varphi_j (x) = \cos (j \pi x)$. Revisiting the boundary conditions, $q_0$ is trivial, and $q_1$ becomes: $q_1(t)=h T(x=1,t) = h \sum_{j=0}^{N} b_j (-1)^j$. Naturally, one has to represent also $w(x,t)$ in the space spanned by $\phi_j (x)$. Once one substituted these expressions into \eqref{AFOU}, multiplied them by the corresponding weight functions and integrated respect to $x$ from $0$ to $1$, one obtains a system of ordinary differential equations (ODE). Here, we exploit that the square of the trial functions $\phi(x)^2$ and $\varphi(x)^2$ are both integrable and after integration they are equal to $1/2$. Since the $\cos$ series have a non-zero part for $j=0$, we handle it separately from the others corresponding to $j>0$.
\begin{itemize}
\item For $j=0$, we have
\begin{align}
\dot b_0 + \partial_x w = 0, \rightarrow \dot b_0 = - h b_0 + q_0
\end{align}
with the upper dot denoting the time derivative, and $a_0=0$ identically.
\item For $j>0$, we obtained
\begin{align}
\dot b_j + {j \pi} a_j&=0, \\
a_j &=\alpha {j \pi} b_j+ \frac{2}{j \pi} (h b_j - q_0),
\end{align}
where the term $2/(j \pi)$ comes from the $\sin$ series expansion of $w$. We note that for $j>0$, $\partial_x w$ does not contribute to the time evolution.
\end{itemize}
Such ODE can be solved easily both numerically and analytically for suitable $q_0$ functions. Figure \ref{fig3} shows the analytical solution programmed in Matlab in order to demonstrate the convergence to the right (physical) solution. In this respect, we refer to \cite{RietEtal18} in which a thorough analysis is presented on the analytical and numerical solution of heat equations beyond Fourier. Our interest is to utilize as less terms as possible of the infinite series, which is able to properly describe the rear side temperature history (i.e., the measured one) from a particular time instant. In other words, we want to simplify the complete solution as much as possible but keeping its physical meaning.

Starting with $j=0$ case, we find that the terms in the particular solution with $\exp(-C_1 t)$ and $\exp(-C_2 t)$ extinct very quickly, thus we can safely neglect them with keeping the $\exp(- h t)$ as the leading term throughout the entire time interval we investigate. Briefly, $j=0$ yields
\begin{align}
b_0(t) = Y_0 \exp(- h t), \quad Y_0=(C_1-C_2)/\big(n(C_1-h) (C_2-h) \big).
\end{align}
Continuing with $j=1$, we make the same simplifications and neglecting the same exponential terms as previously after taking into account the initial condition, and we found
\begin{align}
b_1(t) = Y_1 \exp(- 2h t) \exp(-\pi^2 t), \quad Y_1=2(C_1-C_2)/\big(n(C_1 + x_F) (C_2 + x_F) \big), \quad x_F=-2 h - \alpha\pi^2.
\end{align}
Based on the convergence analysis (Fig.~\ref{fig3}), we suppose that these terms are eligible to properly describe the temperature history after $t>30$ (which is equal to $0.3$ s if $t_p=0.01$ s). Finally, we can combine these solutions, thus $T(x=1,t)=b_0 - b_1$ (the alternating sign originates in $\cos (j \pi)$).

\subsection{Step 2: solving the Guyer-Krumhansl equation} 
\begin{figure}
\centering
\includegraphics[width=13cm,height=10cm]{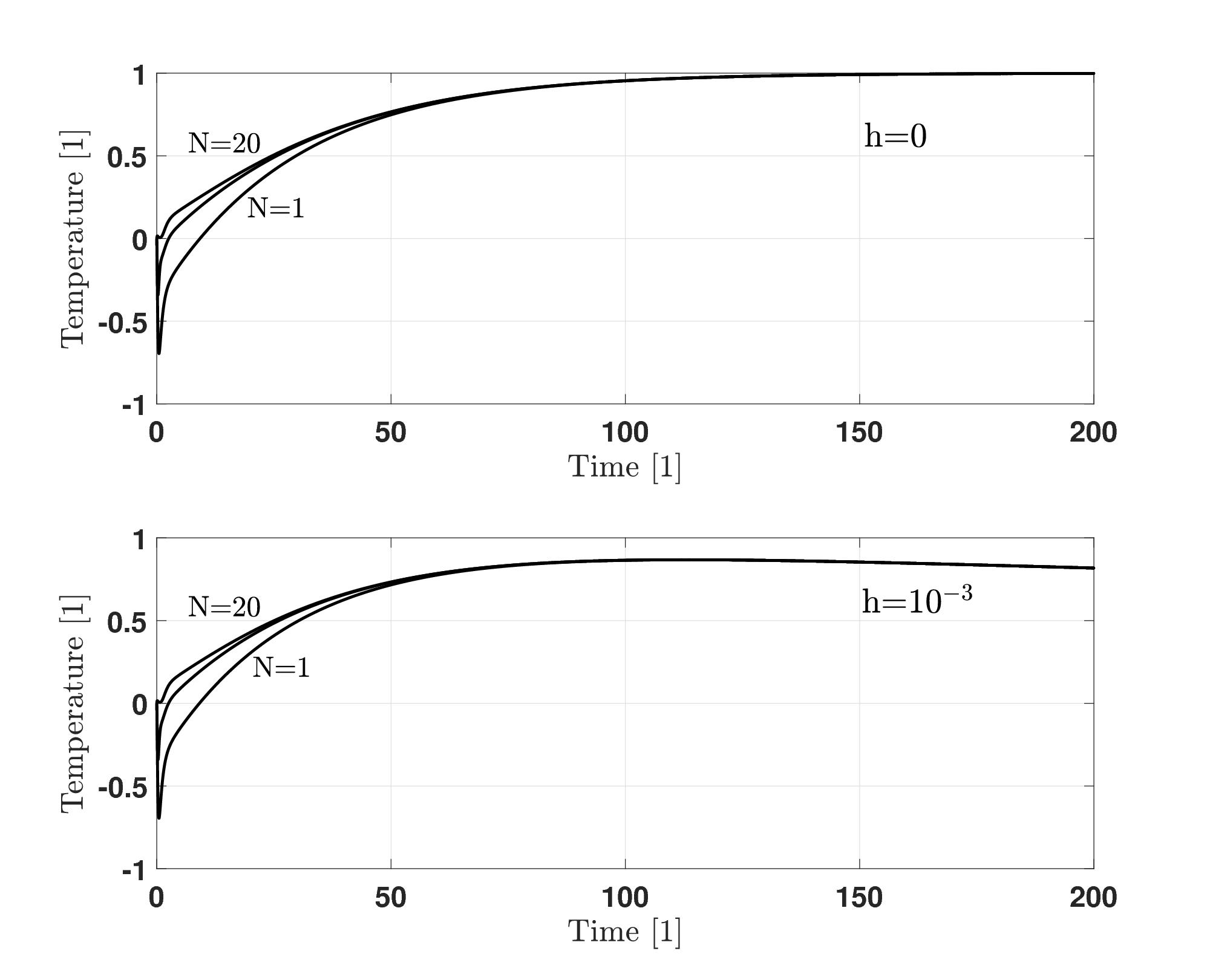}
\caption{Convergence analysis for the Guyer-Krumhansl equation in two different cases on the rear side temperature history. The first one shows the adiabatic limit, and the second one presents the case when the heat transfer coefficient $h$ is not zero. In both cases, we applied $1$, $3$ and $20$ terms in the spectral decomposition \eqref{specFOU}, with $\alpha=0.005$, $\tau_q=1$, and $\kappa^2=10 \alpha \tau_q$.}
 \label{fig4}
\end{figure}

Here, we repeat the calculations using the same set of trial and weight functions for the GK model, that is, we solve
\begin{align}
 \partial_t T + \partial_x q=0, \quad
 \tau_q \partial_t q + q + \alpha \partial_x T - \kappa^2 \partial_{xx} q = 0, \label{AGK}
\end{align}
with
\begin{align}
q_0(t)= -\Big(\exp(-C_1 t/t_p) +\exp(-C_2 t/t_p)\Big)/n, \quad q_1(t) = h T(x=1,t), \quad  q(x,t=0)=0,\quad T(x,t=0)=0.
\end{align}
Analogously with the Fourier case, we obtain a set of ODE as follows.
\begin{itemize}
\item For $j=0$, we have
\begin{align}
\dot b_0 + \partial_x w = 0, \rightarrow \dot b_0 = -h b_0 + q_0,
\end{align}
which is the same as previously due to $a_0=0$ identically.
\item For $j>0$, $a_j$ changes
\begin{align}
\dot b_j + j \pi a_j&=0, \\
\tau_q \dot a_j +\left (1+\kappa^2 j^2 \pi^2\right ) a_j &=\alpha j \pi b_j+ \frac{2}{j \pi} \left [(h b_j - q_0) + \tau_q (h \dot b_j - \dot q_0) \right ].
\end{align}
\end{itemize}
Consequently, the zeroth term, $b_0(t)$ remains the same with the particular solution being omitted,
\begin{align}
b_0(t) = Y_0 \exp(- h t), \quad Y_0=(C_1-C_2)/\big(n(C_1-h) (C_2-h) \big).
\end{align}
However, for $b_1(t)$, the particular solution $P(t)$ becomes more important, its initial value influences the temperature history, i.e., $P_0=P(t=0)$ and $DP_0=\textrm{d}_t P(t=0)$ appears in the coefficients $Z_1$ and $Z_2$, and the $P_0$ and $DP_0$ quantities are important in the evaluation method, too. Thus $b_1(t)$ reads
\begin{align}
b_1(t) = Z_1 \exp{(x_1 t)} + Z_2 \exp{(x_2 t)} + P(t), \quad Z_1=-\frac{DP_0 - P_0 x_2}{x_1-x_2}, \quad Z_2=-P_0 + \frac{DP_0 - P_0 x_2}{x_1-x_2}.
\end{align}
The exponents $x_1$ and $x_2$ depend on the GK parameters $\tau_q$ and $\kappa^2$, and obtained as the roots of the quadratic equation $x_j^2 + k_{1j} x + k_{2j}=0$:
\begin{align}
x_{1,2}=x_{j1,2}|j=1, \quad x_{j1,2}=\frac{1}{2} \left(-k_{1j} \pm \sqrt{k_{1j}^2 - 4 k_{2j}}\right), \quad k_{1j}=\frac{1+ \kappa^2 j^2 \pi^2}{\tau_q} +2 h, \quad k_{2j}=\frac{\alpha j^2 \pi^2}{\tau_q} + \frac{2 h}{\tau_q}.
\end{align}
Furthermore, the particular solution reads as
\begin{align*}
P_j(t) = M_{j1} \exp{(-C_1 t)} + M_{j2} \exp{(-C_2 t)}, \quad M_{j1}=\left(\frac{2 C_1}{n} - \frac{2}{n \tau_q}\right)/\left[k_{2j} - k_{1j} C_1 + C_1^2\right],
\end{align*}
\begin{align}
M_{j2}=\left(-\frac{2 C_2}{n} + \frac{2}{n \tau_q}\right)/\left[k_{2j} - k_{1j} C_2 + C_2^2\right].
\end{align}
Hence $P_0=M_1 + M_2$ and $DP_0 =-M_1 C_1 - M_2 C_2$ appears in $b_1(t)$ with $j=1$, too. After obtaining $Z_1$ and $Z_2$, $P(t)$ can be neglected since it becomes negligibly small at $t>t_p$. Finally, we formulate the rear side temperature history using $b_0(t)$ and $b_1(t)$ as
\begin{align}
 T(x=1,t) =  b_0 -  b_1 =  Y_0 \exp(- h t) -   Z_1 \exp{(x_1 t)} -  Z_2 \exp{(x_2 t)},
\end{align}
for which Figure \ref{fig4} shows the convergence property.

\newpage
\bibliographystyle{unsrt}

\end{document}